\documentclass[12pt]{iopart}

\usepackage{iopams}  
\usepackage{xcolor,graphicx}
\usepackage{amssymb}
\usepackage[utf8]{inputenc}
\usepackage{soul}

\begin{document}

\title[Hamilton-Jacobi hydrodynamics of pulsating relativistic stars]{Hamilton-Jacobi hydrodynamics of pulsating relativistic stars}

\author{John Ryan Westernacher-Schneider}
\address{Department of Astronomy, University of Arizona, Tucson, AZ 85721, USA}
\ead{jwestern@email.arizona.edu}

\author{Charalampos Markakis}
\address{DAMTP, University of Cambridge, Wilberforce Rd, Cambridge CB3 0WA, UK}
\address{Mathematical Sciences, Queen Mary University of London,  Mile End Road London E1 4NS, UK}
\address{NCSA, University of Illinois at Urbana-Champaign, 1205 W Clark St, Urbana, IL 61801, USA}
\ead{c.markakis@damtp.cam.ac.uk}

\author{Bing Jyun Tsao}
\address{Department of Physics, University of Illinois at Urbana-Champaign, 1110 W Green  St, Urbana, IL 61801, USA}

\vspace{10pt}
\begin{indented}
\item[]November 2019
\end{indented}

\begin{abstract}
The dynamics of self-gravitating fluid bodies is described by the Euler-Einstein system of partial differential equations. The break-down of well-posedness on the fluid-vacuum interface remains a challenging open problem, which is manifested in simulations of oscillating or inspiraling binary neutron-stars. We formulate and implement a well-posed canonical hydrodynamic scheme, suitable for neutron-star simulations in numerical general relativity. The scheme uses a variational principle by Carter-Lichnerowicz stating that barotropic fluid motions are conformally geodesic and Helmholtz's third theorem stating that initially irrotational flows remain irrotational. We apply this scheme in 3+1 numerical general relativity to evolve the canonical momentum of a fluid element via the Hamilton-Jacobi equation. We explore a regularization scheme for the Euler equations, that uses a fiducial atmosphere in hydrostatic equilibrium and allows the pressure to vanish, while preserving strong hyperbolicity on the vacuum boundary. The new regularization scheme resolves a larger number of radial oscillation modes compared to standard, non-equilibrium atmosphere treatments.
\end{abstract}

%\maketitle

\section{Introduction}

Gravitational waves from compact binaries carry unique information on their properties
and probe physics inaccessible to terrestrial laboratories \cite{readMeasuringNeutronStar2009,markakisNeutronStarEquation2009,readMatterEffectsBinary2013}. Although development of black-hole gravitational wave templates in the past 15 years has
been revolutionary, the corresponding work for double neutron-star systems has faced challenges, due to the complications in simulating fluids in curved spacetime \cite{Font2008,Rezzolla2013}.
The mathematical description of strongly gravitating fluid bodies requires coupling between the Euler equations of fluid dynamics and the Einstein equations of general relativity. A fundamental open problem is to develop a mathematical framework that establishes existence, uniqueness and global regularity of solutions given some initial conditions, and to track the moving boundary separating the fluid from vacuum, where strong hyperbolicity (and thus well-posedness) of the Euler equation breaks down \cite{christodoulouMotionFreeSurface2000,Lindblad2005}. In a Newtonian context, the degenerate nature of the problem was pointed out by J. von Neumann and W. Heisenberg in 1949 \cite{jangWellposednessCompressibleEuler2009,hadzicPrioriEstimatesSolutions2019}. In a relativistic context, these problems manifest themselves in the hydrodynamic simulation of neutron-star binaries in numerical general relativity, an area that has seen rapid recent developments over the past years \cite{Rezzolla2013,Radice2014}. Ill-posedness on the vacuum boundary prevents stable and meaningful numerical evolution
\cite{hilditchIntroductionWellposednessFreeevolution2013, jangWellposednessCompressibleEuler2015, oliynykLagrangeCoordinatesEinsteinEuler2012, Oliynyk2014, schoepeRevisitingHyperbolicityRelativistic2018, oliynykDynamicalRelativisticLiquid2019}.

Relativistic hydrodynamic  simulations are commonly stabilized via  an artificial atmosphere, but this introduces new artifacts, such as artificial accretion onto the surface, that can prevent point-wise convergence (cf.~Ref.~\cite{Duez2002} for a scheme  that replaces an atmosphere with `if statements'). The resulting error in the mass estimates of ejecta from binary neutron star mergers can  be as high as $\sim 70 \,\%$ \cite{dietrichGravitationalWavesMass2017,theligoscientificcollaborationEstimatingContributionDynamical2017}. However, the artificial atmosphere issue is absent in codes which are not grid-based, e.g.~the smoothed-particle hydrodynamics code of~\cite{stergioulas2011gravitational}.

Arnol'd has described
the nonrelativistic Euler equation as the geodesic equation on the group of volume-preserving diffeomorphisms \cite{Arnold1966}. Synge \cite{Synge37} and Lichnerowicz \cite{Lichnerowicz1967} have shown that the motion of a relativistic  barotropic fluid element can be described as conformally geodesic (cf. Eq.~(\ref{eq:action}) below). Carter  \cite{Carte79}
demonstrated how this  approach leads to  elegant derivations of conservation laws for neutral or charged, poorly conducting fluids, utilizing a super-Hamiltonian form of the Euler equations in 4-dimensional general relativity. Markakis et
al.~\cite{markakisConservationLawsEvolution2017} extended Carter’s framework to perfectly conducting magnetofluids, adopting the
Bekenstein-Oron formulation of ideal magnetohydrodynamics \cite{Bekenstein2000,Bekenstein2001}.
Conservation laws that are Noether-related to helical symmetry  lie at the heart of the self-consistent field method for constructing quasi-equilibrium initial data for 
neutron-star binaries \cite{Gourgoulhon2006,PriceMarkakisFriedman2009,Teukolsky1998,gourgoulhonRelationsThreeFormalisms1998}. Nevertheless, the above framework has not been used for evolving relativistic fluid flows.

To this end, Markakis \cite{markakisHamiltonianHydrodynamicsIrrotational2014}
casted Carter's framework to a form suitable for numerical evolution, using (constrained) Hamiltonian
descriptions of barotropic fluids in Newtonian gravity and 3+1 general relativity.
One may use this Hamiltonian or (for irrotational flows) Hamilton-Jacobi description of fluid dynamics in order to cast the evolution
equations into a hyperbolic form, useful for evolving oscillating, rotating or binary neutron stars in the inspiral phase in
numerical general relativity. The binary inspiral phase is expected to be well-modelled as barotropic~\cite{FriedmanStergioulas2013}, and most binary neutron star simulations start with irrotational initial data, which is considered a good approximation tens of orbits before merger when the orbital frequency is much higher than the spin frequency.
 In the present paper, we implement and test the formulation for single, irrotational, radially oscillating neutron stars in the Cowling approximation, and make comparisons with the Valencia formulation \cite{Font2008}. Due to large perturbations injected into the star from the the stellar surface-vacuum interface, likely due to a different solution structure there in the Hamiltonian formulation~\cite{JRWSbaroclinic}, we found it necessary to use the Valencia formulation near the surface. Our implementation of the Hamiltonian formulation is therefore a hybrid one. Further improvements, possibly using the Hamiltonian formulation everywhere, will be left for future work. In a companion paper, it will be shown that the Hamilton-Jacobi formulation can be super-convergent when evolving slightly perturbed quasi-stationary flows, but has regular convergence when evolving more dynamical flows.

In addition to comparing the Valencia and Hamilton-Jacobi formulations using the standard atmosphere, we will also demonstrate the utility of a fiducial atmosphere treatment which we call the \emph{equilibrium atmosphere}, suitable for either the Hamiltonian or Valencia formulations.
In Sec.~3.5 we will describe a regularization scheme for the Euler equations that maintains 
strong hyperbolicity on the vacuum boundary, where the pressure now vanishes exactly. The scheme can be combined with the equilibrium atmosphere treatment (used here only to impose a reflective boundary condition on the surface; but no longer needed to maintain hyperbolicity), which avoids spurious  accretion or artificial shock heating on the star surface. We demonstrate that this combination results in significantly lower numerical noise in simulations, which allows extraction of higher overtone radial pulsation modes that do not appear using  standard treatments at the same resolution. However, we find the equilibrium atmosphere is significantly more dissipative than the standard atmosphere. Thus, our presentation of the equilibrium atmosphere should be viewed as a preliminary exploration of a novel vacuum regularization technique. Its usefulness in, eg.~binary neutron star simulations, remains to be seen.

We use units in which $G=c=1$ throughout, and the mostly-positive metric signature $(-,+,+,+)$. Spacetime indices are denoted with Greek letters, and spatial indices are denoted with $\lbrace i,j,k...\rbrace$

\section{Hydrodynamic equations of motion}

\subsection{Hamiltonian formulation}

In this section we review the Hamiltonian formulation for a relativistic barotropic perfect fluid with rotation~\cite{markakisHamiltonianHydrodynamicsIrrotational2014}. The energy-momentum tensor reads
\begin{eqnarray} \label{eq:stressenergy}
T^{\mu \nu} = \rho h u^\mu u^\nu + P g^{\mu \nu},
\end{eqnarray}
where
\begin{eqnarray} \label{eq:enthalpydef}
h = 1+\int  \frac{dP}{\rho} = 1+ e +\frac{P}{\rho}
\end{eqnarray}
  is the specific enthalpy, $\rho$ is the rest-mass density, $P$ is pressure,  $e$ is the specific internal energy and $u^\mu=dx^\mu/d\tau$ is the four-velocity of a fluid element. 

For a relativistic fluid with a polytropic equation of state,
\begin{equation} \label{eq:polytrope1}
P = K \rho^\Gamma.
\end{equation}
The polytropic equation of state is a special case of the equation of state for an ideal fluid,
\begin{equation} \label{eq:polytrope2}
P = \rho e (\Gamma -1).
\end{equation}
For the polytropic equation of state, Eq.~(\ref{eq:enthalpydef}) yields
\begin{equation}
h = 1 + \frac{P}{\rho} \left( 1 + \frac{1}{\Gamma -1} \right) %
= 1 + \frac{K \Gamma}{\Gamma -1} \rho^{\Gamma -1}. \label{eq:polyh}
\end{equation}
Note that Eq.~(\ref{eq:polytrope1}) is valid only for barotropic fluids, but Eq.~(\ref{eq:polytrope2}) holds for barotropic as well as baroclinic fluids. Thus,  the latter   can accommodate the entropy generated during shock formation \cite{duezRelativisticMagnetohydrodynamicsDynamical2005}.  In this paper, we focus on  barotropic flows without shocks. Treating shocks within a baroclinic Hamiltonian formulation seems to give unphysical solutions, and thus is not advisable \cite{JRWSbaroclinic}. Therefore, our restriction to the barotropic case is appropriate only to the inspiral phase of a relativistic binary in future applications.

The equations of motion for a barotropic  fluid can be chosen to consist of local rest-mass conservation:
\begin{eqnarray} \label{eq:continuity}
\nabla_\mu \left( \rho u^\mu \right)=\frac{1}{\sqrt{-g}} \partial_\mu(\sqrt{-g} \, \rho u^\mu ) =0,
\end{eqnarray}
which is an approximation of the conservation of baryon number, and the canonical Euler equation:
\begin{equation}
 u^\mu  \left( \partial_\mu p_\nu - \partial_\nu p_\mu \right)=0 \label{eq:hameuler}
\end{equation}
where
\begin{equation}
p_\mu = \frac{\partial L}{\partial u^\mu}=h u_\mu  \label{eq:canonicalmomentum}
\end{equation}
 is the canonical four-momentum of a fluid element, $L$ is the Lagrangian and $g$ is the spacetime metric determinant. 
The canonical Euler equation~(\ref{eq:hameuler})
 follows from  extremizing the action functional
\cite{Lichnerowicz1967,Carte79,markakisHamiltonianHydrodynamicsIrrotational2014,markakisConservationLawsEvolution2017}
\begin{equation}
\mathcal{S} = \int_{\tau_1}^{\tau_2} L(x,u) \, d \tau= \int_{\tau_1}^{\tau_2}  h \sqrt{g_{\mu \nu}\frac{dx^\mu}{d\tau}\frac{dx^\nu}{d\tau}} \,d \tau .\label{eq:action}
\end{equation} 
For barotropic fluids, the canonical Euler equation amounts to covariant conservation of  the energy-momentum tensor
(\ref{eq:stressenergy}),
\begin{equation}
\nabla_\mu T^\mu_{\:\ \nu}=\frac{1}{\sqrt{-g}} \partial_\mu(\sqrt{-g} T^\mu_{\:\ \nu}) -\Gamma^\lambda_{\mu \nu} T^\mu_{\:\ \lambda}=0. \label{eq:Tabconservation}
\end{equation}
In contrast to Eq.~(\ref{eq:Tabconservation}), derivatives appear in an antisymmetric combination in  Eq.~(\ref{eq:hameuler}), which allows one to use partial derivatives even in curved space, without the appearance of metric connection coefficients $\Gamma^\lambda_{\mu \nu}$. 

The temporal component of Eq.~(\ref{eq:hameuler}) is implied by its spatial components, so we may take $\nu=i$ in Eq.~(\ref{eq:hameuler}) without loss of information. To see this, first set $\nu=i$ in Eq.~(\ref{eq:hameuler}) to obtain
\begin{equation}  \label{eq:i_implies_t}
\partial_t p_i -  \partial_i p_t + v^{j}   \omega_{ji}=0,
\end{equation}
where $v^i:=dx^i/dt=u^i/u^t$ is the Eulerian three-velocity of a fluid element measured in local coordinates and $\omega_{ij}= \partial_i p_j - \partial_j p_i$ is the spatial part of the canonical vorticity 2-form. Next, set $\nu=t$ in the left-hand side of Eq.~(\ref{eq:hameuler}) to obtain
\begin{equation}
u^t v^i \left( \partial_i p_t - \partial_t p_i \right) 
=-u^t v^i v^j \omega_{ji} =0\nonumber
\end{equation}
where we used Eq.~(\ref{eq:i_implies_t}) in the first equality and the antisymmetry of $\omega_{ji}$ in the last equality. Thus the $\nu=i$ component of  Eq.~(\ref{eq:hameuler}) implies the $\nu=t$ component.
In nonrelativistic limit, Eq.~(\ref{eq:i_implies_t}) reduces to  the Crocco equation \cite{Gourgoulhon2006}.

\subsection{Hamilton-Jacobi formulation}
In this section we review irrotational case, which permits an alternative formulation in terms of a scalar potential~\cite{markakisHamiltonianHydrodynamicsIrrotational2014}.

For irrotational fluids, the canonical vorticity 2-form $\omega_{\mu \nu}:= \partial_\mu p_\nu - \partial_\nu p_\mu $ vanishes by definition.\ Then, by virtue of the Poincar\'e lemma, the relativistic Euler equation (\ref{eq:hameuler}) is satisfied identically by a closed canonical momentum 1-form
\begin{equation}
p_\mu=\partial_\mu S   \label{eq:pirrotational}
\end{equation}
where $S$ is the velocity potential. By virtue of Helmholtz's third theorem (a corollary to Kelvin's circulation theorem), initially irrotational flows remain irrotational. A remarkable feature of Kelvin's and Helmholtz's theorems is that, since their derivation is independent of the metric \cite{FriedmanStergioulas2013}, they are exact in generic time-dependent spacetimes,  with gravitational waves carrying energy and angular momentum away from a system. Oscillating stars and radiating binaries, if modeled as barotropic fluids with no viscosity or dissipation other than gravitational radiation, exactly conserve circulation. For irrotational initial data, one may thus evolve a Hamilton-Jacobi equation 
\begin{equation}
g^{\mu \nu} \partial_\mu S  \partial_\nu S+h^2=0  \label{eq:HJ4}
\end{equation}
in lieu of the Euler equation~(\ref{eq:hameuler}). Eq.~(\ref{eq:HJ4}) 
was obtained by substituting Eqs.~(\ref{eq:canonicalmomentum}) \& (\ref{eq:pirrotational}) into the constraint
\begin{equation}
g^{\mu \nu} u_\mu u_\nu =-1.  \label{eq:uu}
\end{equation}
Eq.~(\ref{eq:uu}) (and consequently the Hamilton-Jacobi equation (\ref{eq:HJ4})), is a first integral of the Euler equation (\ref{eq:hameuler})  quadratic in the momenta, resulting from the fact that  $g^{\mu \nu}$ is a Killing tensor. This  conserved quantity is Noether-related to the symmetry of the action~(\ref{eq:action}) with respect to proper-time translations, $\tau \rightarrow \tau + \delta \tau$ \cite{Carte79,markakisHamiltonianHydrodynamicsIrrotational2014,markakisConservationLawsEvolution2017}.

With the standard 3+1 decomposition, the spacetime $\mathcal{M}=\mathbb{R} \times \Sigma$ is foliated by a family of spacelike surfaces $\Sigma_t$ and, in a chart $\{ t,x^i\}$, its metric takes the form
\begin{equation}
ds^2 = g_{\mu \nu} dx^\mu dx^\nu=
-\alpha^2 dt^2+\gamma_{ij}(dx^i+\beta^i dt)(dx^j+\beta^j dt) \label{eq:31metric}
\end{equation}
where $\alpha$ is the lapse, $\beta^a$ is the shift vector and $\gamma_{ab}$ is the spatial metric. Substituting the 3+1 metric into the Hamilton-Jacobi equation~(\ref{eq:HJ4}) yields a quadratic equation for $\partial_t S$. Of the two algebraic roots, the one with the correct Newtonian limit \cite{markakisHamiltonianHydrodynamicsIrrotational2014} is:
\begin{equation} \label{eq:HJ31irr}
%\partial_t S-\beta^i \partial_i S+\alpha \sqrt{\gamma^{ij} \partial_i S\partial_jS+h^2}=0
{\partial _t}S\underbrace { - {\beta ^i}{\partial _i}S + \alpha \sqrt {{\gamma ^{ij}}{\partial _i}S{\partial _j}S + {h^2}} }_H = 0
\end{equation}
or, equivalently,
\begin{equation}
\partial_t S+H=0  \label{eq:HJ31}
\end{equation}
where 
\begin{eqnarray}
H&=&-\beta^i p_i +\alpha \sqrt{\gamma^{ij} p_i p_j+h^2} \label{eq:Hamiltonian} \\
p_i&=&\partial_i S \label{eq:canmomirr}
\end{eqnarray}
Eqs.~(\ref{eq:HJ31})--(\ref{eq:canmomirr}) amount to 3+1 decompositions of Eqs.~(\ref{eq:pirrotational})--(\ref{eq:HJ4}). 
Solutions to Hamilton-Jacobi equations are non-unique, albeit viscosity solutions are unique \cite{Gomes2002,Bressan2003}. One option in a numerical scheme would be to evolve Eq.~(\ref{eq:HJ31})  directly, which would guarantee the irrotationality of the flow since the canonical momentum would be computed as the gradient of the scalar potential. Such an approach is left to future work. In this work, we instead opt to solve for the gradient of the Hamilton-Jacobi equation~(\ref{eq:HJ31}), in the form  of a hyperbolic conservation law \cite{jinLevelSetMethod2003}:
\begin{equation}
\partial_t p_i+\partial_iH=0,  \label{eq:HJgrad31}
\end{equation}
subject to the constraint
\begin{equation}
\partial_i p_j - \partial_i p_j=0.  \label{eq:HJconstraint}
\end{equation}
This simply amounts to setting $\omega_{ji}=0$ in Eq.~(\ref{eq:i_implies_t}). As shown in \cite{markakisHamiltonianHydrodynamicsIrrotational2014,markakisConservationLawsEvolution2017}, the constrained Hamiltonian~(\ref{eq:Hamiltonian}) is opposite to the  time component of the  canonical four-momentum~(\ref{eq:canonicalmomentum}), 
\begin{equation}
H=-p_t.  \label{eq:Hpt}
\end{equation}
Thus, for vanishing vorticity,  Eq.~(\ref{eq:HJgrad31}) is equivalent to  the canonical Euler equation~(\ref{eq:i_implies_t}). 
 
We note that  Eq.~(\ref{eq:HJgrad31}) is exactly flux-conservative, although no symmetry assumptions about the gravitational field were made.
While helical symmetry (circularized orbits due  to gravitational radiation)
is typically assumed along with irrotationality (negligible spin frequency
compared to the orbital frequency) when constructing initial data in late
inspiral
\cite{Gourgoulhon2006,PriceMarkakisFriedman2009,Teukolsky1998,gourgoulhonRelationsThreeFormalisms1998},
the two assumptions are  independent. Indeed, irrotational (or spinning)
binaries on eccentric (or circular) orbits have  been constructed and evolved
by  Moldenhauer et al.
\cite{Moldenhauer2014,dietrichBinaryNeutronStars2015}. 
Here, we do not assume existence of a  Killing vector field (helical or
otherwise); our sole assumption is that the initial data is irrotational.
Helmholtz's third theorem then guarantees that the data will remain irrotational
throughout the inspiral.

The fact that $H(x,p)$ is  the constrained Hamiltonian of a fluid element can be confirmed by rewriting the action integral~(\ref{eq:action}) using coordinate time $t$ as integration variable, and performing a Legendre transform on the Lagrangian, which yields \cite{markakisHamiltonianHydrodynamicsIrrotational2014}
\begin{equation}
S = \int_{t_1}^{t_2} [v^i p_i-H(x,p)] \, d t
= \int_{t_1}^{t_2} [v^i p_i+\beta^i p_i -\alpha \sqrt{\gamma^{ij} p_i p_j+h^2}] \, d t.
\label{eq:action2}
\end{equation}

It is common to introduce the Lorentz factor
\begin{equation}
W = \alpha u^t=\frac{1}{\sqrt{1-\gamma_{ij}\nu^i \nu^j}} \\ \label{eq:LorentzW}
\end{equation}
where $\nu^i=\alpha^{-1} (v^i+\beta^i)=\alpha^{-1} (u^i/u^t+\beta^i)$ is the fluid  three-velocity measured by normal observers. Eq.~(\ref{eq:Hpt}) implies that the constrained Hamiltonian~(\ref{eq:Hamiltonian}) can be written as
\begin{equation}
H=hW  \left(\alpha- \gamma_{ij}\nu ^i\beta^j \right) \label{eq:HamLorentzW}
\end{equation}
where we used the 3+1 metric~(\ref{eq:31metric}) to lower the indices in  $p_t=g_{t\mu}p^\mu$. Similarly, using the 3+1 metric to lower the indices in $p_i=g_{i\mu}p^\mu$, the spatial components of the canonical momentum~(\ref{eq:canonicalmomentum}) can be written as 
\begin{equation}
p_i=hW\gamma_{ij}\nu ^j
\end{equation}
Then, the Hamilton-Jacobi conservation law~(\ref{eq:HJgrad31}) takes the form
\begin{equation}
\partial_t (hW\gamma_{ij}\nu ^j)+\partial_i [hW  \left(\alpha- \gamma_{ij}\nu ^i\beta^j \right)]=0.  \label{eq:HJgrad3131}
\end{equation}
The spacetime metric determinant $g$ is related to the spatial metric determinant $\gamma$ via $\sqrt{-g} = \alpha \sqrt{\gamma}$, where $\alpha$ is the lapse function. The rest-mass conservation law~(\ref{eq:continuity}) can then be written as
\begin{equation}
 \partial_t \left( \sqrt{\gamma} \rho W \right) + \partial_i \left[ \alpha \sqrt{\gamma}\rho W \left(\nu^i - \beta^i/\alpha \right) \right]=0. \label{eq:continuity31}
\end{equation}
This Hamilton-Jacobi formulation for the barotropic fluid therefore consists of Eqs.~(\ref{eq:HJgrad3131})-(\ref{eq:continuity31}), with the equation of state $h=h(\rho)$ given by Eq.~(\ref{eq:enthalpydef}) or, for a polytrope, Eq.~(\ref{eq:polyh}). 
Coupling with gravity enters through the components of the metric~(\ref{eq:31metric}),
which satisfies the Einstein equations. Note that Eqs.~(\ref{eq:HJgrad3131})-(\ref{eq:continuity31}) are source-free in arbitrary dimensions. In this paper, we will consider  the spacetime metric fixed, and evolve the hydrodynamic equations only (Cowling approximation).

In the Newtonian limit, the Hamilton-Jacobi equation~(\ref{eq:HJ31irr}) reduces to
\begin{equation} \label{eq:HJ31irrNewt}
%\partial_t S+\frac{1}{2}\gamma^{ij} \partial_i S\partial_jS+h+\Phi=0
{\partial _t}S + \underbrace {\frac{1}{2}{\gamma ^{ij}}{\partial _i}S{\partial _j}S + h + \Phi }_H = 0
\end{equation}
where $\Phi$ is the Newtonian gravitational potential \cite{chenMathematicsShockReflectionDiffraction2018}. 
This equation has typically been obtained as a first integral to the nonrelativistic, irrotational Euler equation and has sometimes been  referred to  as 
a  ``Bernoulli-type theorem'' for  non-steady irrotational flows\footnote{The original Bernoulli theorem is a conservation law along streamlines only and is Noether-related to a Killing symmetry. The first integral~(\ref{eq:HJ31}) of Eq.~(\ref{eq:HJgrad31}) is constant throughout the fluid, and amounts   simply to the Hamilton-Jacobi equation. Eq.~(\ref{eq:HJ31}) and its gradient,  the conservation law (\ref{eq:HJgrad31}),  hold for all irrotational flows without symmetry assumptions.};  the function $H=\frac{1}{2}v^2+h+\Phi$ has been referred to as the ``Bernoulli function'' or the ``total head'' in engineering literature. 
Blandford and Thorne \cite{blandfordrogerModernClassicalPhysics2017} use the more physically motivated term 
``injection energy'' (the energy required to bring  a fluid element from infinity and inject it  into a self-gravitating  fluid with the same chemical potential and velocity as the surrounding elements). In light of
the above discussion, we will refer to this function simply as the Hamiltonian of a fluid element (which coincides with the energy of a fluid element measured in local coordinates).

\subsection{Specialization to Minkowski space in $1+1$ dimensions.}
In $1$ spatial dimension, the canonical vorticity vanishes identically, and
Eq.~(\ref{eq:i_implies_t}) takes   the flux-conservative form of the Hamilton-Jacobi equation~(\ref{eq:HJgrad3131}).  In flat spacetime in Cartesian coordinates, we have
\begin{equation}
\partial_t \left( h W v^x \right) + \partial_x \left( h W \right)=0 \label{eq:hampolyeuler1D}
\end{equation}
where we set $\gamma_{ij}=\delta_{ij}, \alpha=1, \beta^i=0$ and
 $ W = 1/\sqrt{1-(v^x)^2 }$ is the Lorentz factor. Similarly, the continuity equation~(\ref{eq:continuity31}) becomes 
\begin{equation}
\partial_t \left( \rho W \right) + \partial_x \left(  \rho W v^x \right)=0. \label{eq:restmass1D}
\end{equation}

In the case of a polytropic fluid, the Hamilton-Jacobi formulation   consists of Eqs.~(\ref{eq:hampolyeuler1D})--(\ref{eq:restmass1D}), with the specific enthalpy $h(\rho)$ given by Eq.~(\ref{eq:polyh}). The conservative variables are $D \equiv \rho W$ and $ p_x = p^x =h W v^x$.

We contrast this with the analogous Valencia formulation, where in lieu of Eq.~(\ref{eq:hampolyeuler1D}) we have Eq.~(\ref{eq:Tabconservation}) arising from energy-momentum conservation,
\begin{equation}
\partial_t \left( \rho h W^2 v^x \right) + \partial_x \left( \rho h W^2 v^x v^x + K \rho^\Gamma \right)=0,
\end{equation}
where we substituted $u_i = u^i = W v^i$ into Eq.~(\ref{eq:stressenergy}). 

 The primitive variables of the Hamilton-Jacobi formulation can be recovered from the conservative variables by root-finding for $\rho$ on the expression
\begin{equation}
(p^x)^2 \rho^2 = h(\rho)^2 \left( D^2 - \rho^2 \right),
\end{equation}
which is obtained by writing $(p^x)^2 = h^2 W^2 (v^x)^2$ and then using $(v^x)^2 = (W^2-1)/W^2 $ and $W = D/\rho$. Once $\rho$ is recovered, the Lorentz factor is obtained via $W=D/\rho$ and then the velocity is recovered via $v^x = p^x/(h(\rho) W)$.

In the case of dust, that is, a zero pressure fluid, one has $h=1$ and Eq.~(\ref{eq:hampolyeuler1D}) becomes
\begin{equation}
\partial_t \left(W v^x \right) + \partial_x W =0 \label{eq:hampolyeuler1D_dust}
\end{equation}
This is a relativistic generalization of the inviscid Burgers equation  \cite{LeFloch2012a}, which  is recovered in the non-relativistic limit, $ v^x \ll1$, whence $W v^x  \simeq  v^x$ and $ W \simeq 1+\frac{1}{2}(v^x)^2 $.
\subsection{Specialization to curved space in spherical symmetry}
In curved spacetime, in spherical symmetry, the metric can be written as
\begin{eqnarray}
ds^2 = \left(-\alpha^2 + \gamma_{rr}\beta^r\beta^r \right) dt^2 + 2 \gamma_{rr}\beta^r dt dr + \gamma_{rr} dr^2 + \gamma_{T}r^2 d\Omega^2,
\end{eqnarray}
where $d\Omega^2 = d\theta^2 + \sin^2\theta d\phi^2$. The metric functions $\alpha$, $\beta^r$, $\gamma_{rr}$, and $\gamma_T$ are functions of $(t,r)$ only. We then have $\sqrt{-g} = \alpha r^2 \sqrt{\gamma_{rr}} \gamma_T$ (we set $\theta = \pi/2$ so that $\sin \theta = 1$, a choice which is permitted in spherical symmetry). Other relations we require are $u^r =u^t v^r= u^t (\alpha \nu^r - \beta^r)$, $W = \alpha u^t$, and $u_r = W \nu_r$, where $\nu^r$ is the  radial  velocity of a fluid element measured by normal observers and $v^r=dr/dt$ is  the radial velocity of a fluid element measured in local coordinates.

With this preamble in hand, the  continuity equation~(\ref{eq:continuity31})  becomes
\begin{eqnarray}
\partial_t \tilde{D} + \frac{1}{r^2} \partial_r \left[ \alpha r^2 \tilde{D} \left(\nu^r - \beta^r/\alpha \right) \right]=0, \label{eq:restmassSphSym}
\end{eqnarray}
where $D \equiv \rho W$ and the tilde denotes densitization with $\sqrt{\gamma_{rr}}\gamma_T$, i.e. $\tilde{D} = \sqrt{\gamma_{rr}}\gamma_T D$.
The derivatives in the Euler equation in the Hamilton-Jacobi form~(\ref{eq:HJgrad3131}) appear in an antisymmetric combination, which allows one to forgo the introduction of metric determinants. It is simply
\begin{eqnarray}
\partial_t p_r + \partial_r H=\partial_t (hW\gamma_{rr}\nu ^r)+\partial_r [hW  \left(\alpha- \gamma_{rr}\nu ^r\beta^r \right)]=0. \label{eq:HamEulerSphSym}
\end{eqnarray}
The coupling with the metric enters through the relation between the contravariant and covariant forms of the canonical momentum. Namely, $H=-p_t = -g_{t\mu} p^\mu = hW (\alpha - \gamma_{rr} \beta^r \nu^r)$ and $p_r = h u_r = hW\gamma_{rr} \nu^r$.

The system of hydrodynamic equations on a curved spherically symmetry background for the barotropic fluid therefore consist of an equation of state $h = h(\rho)$ (in particular, Eq.~(\ref{eq:polyh}) for a polytrope), and Eqs.~(\ref{eq:restmassSphSym}) \&~(\ref{eq:HamEulerSphSym}). Notice that no geometric source terms appear in this canonical system of equations, in contrast with  equation~(\ref{eq:Tabconservation}) stemming from   energy-momentum conservation. Geometric source terms were not able to be eliminated in a different class of formulations considered in~\cite{papadopoulos1999analysis}.

\subsubsection{Primitive variable recovery.}
Similarly to the Minkowski case, one can recover the primitive variables $(\rho,\nu^r)$ from the convervative variables $(\tilde{D},p_r)$ by first root-finding for $\rho$ on the function $f(\rho)$ defined by
\begin{eqnarray}
f(\rho) = -\rho^2 p_r^2 + h(\rho)^2 \gamma_{rr} \left( D^2 - \rho^2 \right). \label{eq:rootfind}
\end{eqnarray}
Note that the relativistic density $\tilde{D}$ has been undensitized in this formula, i.e. $D = \tilde{D}/(\sqrt{\gamma_{rr}} \gamma_T)$. After obtaining $\rho$, we compute $W = D/\rho$ and then recover the velocity via $\nu^r = p_r/(\gamma_{rr} h(\rho) W)$.

One could just as well solve for the enthalpy $h$, which would avoid division by $\rho$.

\subsubsection{Characteristic structure.} \label{sec:char}

One potentially awkward issue for studying the characteristic structure of the spherically symmetric system is the presence of a $1/r^2$ prefactor in the flux term of Eq.~(\ref{eq:restmassSphSym}) and simultaneously its absence in Eq.~(\ref{eq:HamEulerSphSym}). We reason as follows.

By expanding the flux term, the rest mass conservation equation can be written as
\begin{eqnarray}
0 &=& \partial_t \tilde{D} + \partial_r \left[\alpha \tilde{D} \left( \nu^r - \beta^r/\alpha \right)\right] + \frac{2}{r} \alpha \tilde{D} \left( \nu^r - \beta^r/\alpha \right).
\end{eqnarray}
The last term is lower order in the sense of characteristic analysis. Thus, for the purposes of computing the characteristic structure, the equations of motion for this system can be cast as
\begin{eqnarray}
0 &=& \partial_t \vec{U} + \partial_r \vec{F} + \rm{ l.o.},
\end{eqnarray}
where $\vec{U} = (\tilde{D}, p_r)$ are the conservative variables, the fluxes are $\vec{F} = (\alpha \tilde{D} (\nu^r - \beta^r/\alpha), \alpha hW)$, and $\rm {l.o.}$ stands for lower order terms (that is, the non-principal part, in the sense of characteristic analysis).

Our task is to compute the Jacobian $\partial \vec{F}/\partial \vec{U}$. Taking derivatives of the fluxes with respect to the conservative variables is potentially complicated. We can instead take derivatives with respect to primitive variables as follows. Define $\vec{q} = (\rho, \nu^r)$ to be the primitive variables. Then we can use the chain rule to write
\begin{eqnarray}
\frac{\partial \vec{F}}{\partial \vec{U}} &=& \frac{\partial \vec{F}}{\partial \vec{q}} \cdot \frac{\partial \vec{q}}{\partial \vec{U}} 
= \frac{\partial \vec{F}}{\partial \vec{q}} \cdot \left[ \frac{\partial \vec{U}}{\partial \vec{q}} \right]^{-1}. \label{eq:easyJac}
\end{eqnarray}
In this way, we can compute the Jacobian of the system by taking only derivatives of $\vec{F}$ and $\vec{U}$ with respect to primitive variables, which is an easy task, and then performing a matrix inverse of $\partial \vec{U}/\partial \vec{q}$ and multiplying on the left by $\partial \vec{F}/ \partial \vec{q}$.

Define $\Upsilon \equiv \sqrt{\gamma_{rr}}\gamma_T$ for clarity of notation in what follows. We have
\begin{equation}
\frac{{\partial \vec U}}{{\partial \vec q}}{\rm{ }} = \left[ {\begin{array}{*{20}{c}}
{\frac{{\partial \tilde D}}{{\partial {\rho _0}}}}&{\frac{{\partial \tilde D}}{{\partial {\nu^r}}}}\\
{\frac{{\partial {p_r}}}{{\partial {\rho _0}}}}&{\frac{{\partial {p_r}}}{{\partial {\nu^r}}}}
\end{array}} \right] = \left[ {\begin{array}{*{20}{c}}
{\Upsilon W}&{\Upsilon \rho {W^3}{\gamma _{rr}}{\nu^r}}\\
{({\partial _\rho }h)W{\gamma _{rr}}{\nu^r}}&{h{\gamma _{rr}}W(1 + {W^2}{\gamma _{rr}}{\nu^r}{\nu^r})}
\end{array}} \right]
\label{eq:dUdq}
\end{equation}
Note that we have written this for a general barotropic equation of state $h=h(\rho)$. For the polytrope case, substitute $\partial_{\rho} h =  K \Gamma \rho^{\Gamma-2}$. Next, we have
\begin{equation}
\frac{{\partial \vec F}}{{\partial \vec q}}{\rm{ }} = \left[ {\begin{array}{*{20}{c}}
{\Upsilon \alpha W({\nu^r} - {\beta ^r}/\alpha )}&{\Upsilon \alpha \rho W[1 + {W^2}{\gamma _{rr}}{\nu ^r}({\nu^r} - {\beta ^r}/\alpha )]}\\
{({\partial _\rho }h)W(\alpha  - {\gamma _{rr}}{\beta ^r}{\nu^r})}&{hW{\gamma _{rr}}[{W^2}{\nu^r}(\alpha  - {\gamma _{rr}}{\beta ^r}{\nu^r}) - {\beta ^r}]}
\end{array}} \right]
\label{eq:dFdq}
\end{equation}

In our numerical implementation, at each point of the grid we obtain the characteristic structure by computing Eqs.~(\ref{eq:dUdq})~\&~(\ref{eq:dFdq}) and then use Eq.~(\ref{eq:easyJac}) to obtain the Jacobian matrix. We then extract the eigenvalues and, if needed, the left and right eigenvectors using numerical algorithms \cite{NumpyNumpy2019}. In spherical symmetry, this numerical overhead is acceptable, however in a higher dimensional application one ought to compute and simplify analytic formulas for the eigenvalues and eigenvectors.

We give the analogous matrices to Eqs.~(\ref{eq:dUdq}) \&~(\ref{eq:dFdq}) for the Valencia formulation in~\ref{app:valencia}.

\section{Numerical implementation}
In this section we provide details of our numerical implementation.

\subsection{Discretization}
We use a finite volume approach. Let the integer $i$ denote uniformly spaced cell centers and half integers (eg. $i+1/2$) denote cell interfaces, and let $n$ denote the time level. We regulate the $(1/r^2)\partial_r$ term in Eq.~(\ref{eq:restmassSphSym}) in the standard way by replacing it with $3\partial_{r^3}$, which is equivalent by the chain rule. Define the flux vector from Sec.~(\ref{sec:char}) as $\vec{F} = (F_{\tilde{D}}, F_{p_r})$ for brevity. Then the discretized form of Eqs.~(\ref{eq:restmassSphSym}) \&~(\ref{eq:HamEulerSphSym}) are
\begin{eqnarray}
 (\tilde{D})^{n+1}_i  &=& (\tilde{D})^n_i - 3 \frac{\Delta t}{\Delta(r^3)_i} \left[ r_{i+1/2}^2 ({F_{\tilde{D}}})^{n+1/2}_{i+1/2} - r_{i-1/2}^2 ({F_{\tilde{D}}})^{n+1/2}_{i-1/2} \right] \\
 ({p_r})_{i}^{n+1}  &=& ({p_r})_{i}^n - \frac{\Delta t}{\Delta r} \left[ ({F_{p_r}})^{n+1/2}_{i+1/2} - ({F_{p_r}})^{n+1/2}_{i-1/2} \right].
\end{eqnarray}
The subscripts denote spatial positions, and superscripts denote times. We also define $\Delta(r^3)_i \equiv (r_i+\Delta r/2)^3 - (r_i - \Delta r/2)^3$ and $r_{i\pm 1/2} \equiv r_i \pm \Delta r/2$.

\subsection{HLL flux}
We approximate the fluxes at the half time step using the Harten, Lax and van Leer (HLL) formula,
\begin{eqnarray}
F^{\mathrm{HLL}} = \frac{s_R F_L - s_L F_R + s_L s_R \left(U_R - U_L \right)}{s_R-s_L}. \label{eq:HLL}
\end{eqnarray}
Here, $U_R$ and $U_L$ are the conservative variables at the cell interfaces built out of primitive variables which have been reconstructed from their cell-centered values to the right and left of the cell interface, respectively, using the minmod slope limiter. That is,
\begin{eqnarray}
{U_L}_{,i+1/2} &=& U_{i} + \frac{1}{2} \mathrm{minmod}\left(U_{i+1}-U_i,U_i-U_{i-1}\right) \\
{U_R}_{,i+1/2} &=& U_{i+1} - \frac{1}{2} \mathrm{minmod}\left(U_{i+2}-U_{i+1},U_{i+1}-U_{i}\right).
\end{eqnarray}
The scalars $s_R$ and $s_L$ represent the fastest right- and left-moving characteristic speeds among the $U_R$ and $U_L$ states, i.e.
\begin{eqnarray}
s_R &=& \mathrm{max}\left(0, \mathrm{max}\left( \left\{\lambda_R\right\} \right), \mathrm{max}\left( \left\{\lambda_L\right\} \right)\right) \\
s_L &=& \mathrm{min}\left(0, \mathrm{min}\left( \left\{\lambda_R\right\} \right), \mathrm{min}\left( \left\{\lambda_L\right\} \right)\right),
\end{eqnarray}
where $\left\{ \lambda_R \right\}$ and $\left\{ \lambda_L \right\}$ represent the set of all eigenvalues of $(\partial \vec{F}/\partial \vec{U})\vert_{U_R}$ and $(\partial \vec{F}/\partial \vec{U})\vert_{U_L}$, respectively.

\subsection{Comparison of formulations}
We compare two formulations: Valencia, and a hybrid of Valencia and the Hamiltonian formulation. The hybrid formulation uses the Hamiltonian formulation at all points interior to a specified grid point $i_{\mathrm{mix}}$, and Valencia at all points exterior to and including $i_{\mathrm{mix}}$. We find the hybrid scheme is necessary to stabilize the stellar surface, with $i_{\mathrm{mix}}$ chosen to be inside the star so that Valencia is used at the surface. We find that stabilization of the stellar surface is achieved even with the extreme choice of $i_{\mathrm{mix}}$ = $i_{\mathrm{surface}}$, where $i_{\mathrm{surface}}$ is the last interior point of the star. However, more stabilization is achieved with $i_{\mathrm{mix}} \leq i_{\mathrm{surface}}-1$, with minimal differences within that range. Using the Hamiltonian formulation around the stellar surface results in significantly more fluctuations injected into the star by the vacuum regularization routines, which is an issue we will explore in more depth in future work.

\subsection{Ancillary code details}
Here we describe some additional relevant details of our numerical implementation.

We use a 3rd-order total variation diminishing Runge-Kutta time integrator \cite{gottliebTotalVariationDiminishing1998}. For a system of equations of motion written schematically as $\partial_t U = \mathcal{L}(U)$ where $\mathcal{L}$ is a spatial differential operator, the update is described sequentially by
\begin{eqnarray}
U_1 &=& U^n + \Delta t \mathcal{L}\left(U^n\right) \\
U_2 &=& \frac{3}{4} U^n + \frac{1}{4} \left( U_1 + \Delta t \mathcal{L}\left( U_1 \right)\right) \\
U^{n+1} &=& \frac{1}{3} U^n + \frac{2}{3} \left( U_2 + \Delta t \mathcal{L} \left(U_2\right) \right).
\end{eqnarray}

As is standard practice for grid-based computational fluid dynamics, we regularize vacuum regions by imposing an artificial atmosphere there. This amounts to defining a ``floor" value of the rest mass density $\rho_{\mathrm{floor}}> 0$. Whenever primitive variables are computed, either after a conservative-to-primitive variable transformation or after reconstructing the primitive variables at the cell interfaces, we impose a minimum value on $\rho$ given by $\rho_{\mathrm{floor}}$. In the conservative-to-primitive variable transformation routine, the conserved density $D$ is prepared at the outset by the enforcement of $D> \rho_{\mathrm{floor}}$. This bound is implied by $\rho > \rho_{\mathrm{floor}}$. If $\rho$ (or $D$) is found to have a value below $\rho_{\mathrm{floor}}$, then we set $\rho = \rho_{\mathrm{floor}}$ and $\nu^r =0$ (or $D = \rho_{\mathrm{floor}}$ and $p_r = 0$). During evolution, these adjustments tend to be necessary at stellar surfaces, for example. We also impose a maximum speed $\sqrt{\gamma_{rr} \nu^r \nu^r}< \nu_{\mathrm{max}} = 0.99$, although this is never invoked in the evolutions we present in this work. With this atmosphere treatment (herein referred to as the \emph{standard atmosphere}), we find less noisy evolution at high resolution if we \emph{do not} recompute the conservative variables following these atmosphere adjustments, and so this is what we do. This amounts to imposing the vacuum regularization only on the fluxes. This is due to the location of the outer boundary; if placed farther away, recomputing the conservative variables does not introduce noise. Our standard setting is $\rho_{\mathrm{floor}} = 10^{-13}$. This is to be compared with the central density of our TOV star $\rho \vert_{r=0} \equiv \rho_c = 1.28\times 10^{-3}$ in code units.

When rootfinding on Eq.~(\ref{eq:rootfind}), we use Brent's method with hyperbolic extrapolation \cite{ScipyScipy2019}. This is a bracketing method, which therefore requires an initial bracket of the root. We guess the initial bracket to be $[(1.1)^{-1}\times\rho_{\mathrm{floor}}, 1.1\times D]$. If this guess does not bracket the root, then new bracket guesses are generated by widening the initial guess. If this procedure fails to generate a valid bracket, the code aborts. For the evolutions in this work we find the initial bracket guess to be adequate. We set the rootfinder's absolute and relative tolerance parameters to $10^{-60}$ and $10^{-15}$, respectively. In practice, in our simulations this means recovering the primitive variables close to machine precision. We establish this by taking primitive variable snapshots from our simulations, computing the corresponding conservative variables, then passing those conservative variables through the conservative-to-primitive recovery routine and comparing the result with what we started with.

We impose reflecting boundary conditions at $r=0$, namely $p_r|_{r=0} = 0$ by odd parity, and we use an even-parity extrapolation from $r=\{\Delta r, 2\Delta r\}$ for $\tilde{D}$: $\tilde{D}|_{r=0} = (4 \tilde{D}|_{r=\Delta r} - \tilde{D}|_{r=2\Delta r})/3$. At the outer boundary $r=12$ (in code units), we freeze the variables to $p_r = 0$ and $\tilde{D} = \rho_{\mathrm{floor}}$. 

The spacetime metric is a fixed TOV solution in Schwarzschild-like coordinates, thus the evolutions are performed in the Cowling approximation.

\subsection{Regularization of the Euler equations on the vacuum boundary: equilibrium atmosphere}
Simulations of oscillating or binary neutron stars in numerical general relativity have almost always incorporated an artificial atmosphere, to address issues that arise on the stellar surface. The reasons an atmosphere is needed include:

\begin{enumerate}
\item 
The Valencia formulation requires division of $T^t_i=\rho h u^t u_i$ by $\rho u^t$  to recover the variable  $h u_i$ and the primitive variables in each time step. The density vanishes on the stellar surface, where division by zero occurs. An atmosphere keeps the density $\rho$ positive everywhere and avoids division by zero on the stellar surface.
\item
When the sound speed vanishes, the Euler equations become ill-posed due to loss of strong hyperbolicity at the vacuum boundary \cite{jangWellposednessCompressibleEuler2009,hadzicPrioriEstimatesSolutions2019,jangWellposednessCompressibleEuler2015,oliynykLagrangeCoordinatesEinsteinEuler2012,Oliynyk2014,schoepeRevisitingHyperbolicityRelativistic2018,oliynykDynamicalRelativisticLiquid2019}. Well-posedness is maintained via an atmosphere which keeps the density, pressure and thus the sound speed, strictly positive.
\item
Boundary conditions on the neutron-star surface must be that of a free surface in order to obtain the correct oscillation modes~\cite{FriedmanStergioulas2013}. Past studies of fluid-vacuum interfaces suggest that this behavior is recovered in the limit as the atmosphere density tends toward zero~\cite{toroRiemannSolversNumerical2009, munz1994tracking}.
\end{enumerate}

The Hamiltonian formulation avoids issue (i) above, as it directly evolves the variable $p_i=hu_i$.  Since $h=1$ on the stellar surface, no division by zero occurs. However, it does not avoid issues (ii) and (iii). In this work, we introduce an equation of state regularization scheme that keeps the density and sound speed positive when the pressure vanishes (i.e. on the stellar surface). This will eliminate issues (i) and (ii) from the above list. Thus, an atmosphere is no longer required to maintain strong hyperbolicity, and  we are moreover able to reach zero pressure on the stellar surface.
We will still use a fiducial atmosphere in order to impose reflective boundary condition and obtain the correct mode frequencies, per reason (iii) above.

In particular, we  demonstrate the utility of a fiducial atmosphere treatment which we call the \emph{equilibrium atmosphere}. The benefits of this alternative treatment also extend to both the Valencia and the Hamiltonian formulation, but are easier to understand in the latter. The basic idea of the equilibrium atmosphere is to use an equation of state on the entire domain (including the star) which yields a constant Hamiltonian everywhere. A constant Hamiltonian (and zero velocity) implies an equilibrium configuration (see Eqs.~(\ref{eq:HamEulerSphSym}) and~(\ref{eq:restmassSphSym})).

In order for the $\nu^r=0$ Hamiltonian $H= \alpha h$ to remain constant beyond the stellar radius, we must allow the specific enthalpy $h$ to become less than 1. We will achieve this in such a way that the pressure becomes negative, but the rest-mass density stays positive. We will use two generalized polytropes attached piecewise at $h=1$,
\begin{eqnarray}
\rho (h) = \left\{ {\begin{array}{*{20}{l}}
{{{\left( {\frac{{h - 1 + a}}{{K(1 + n)}}} \right)}^n},}&{h > 1}\\
{{{\left( {\frac{{h - 1 + {a^\prime }}}{{{K^\prime }(1 + {n^\prime })}}} \right)}^{{n^\prime }}},}&{h \le 1}
\end{array}} \right.
\label{eq:genpoly}
\end{eqnarray}
The constant $a\ll 1$ is a small regularization parameter that keeps the rest-mass density and the sound speed finite when $h=1$ (or $P=0$). This addresses points (i) and (ii) above, as it avoids the division by zero (in the Valencia formulation) and retains the strong hyperbolicity of the Euler equations on the stellar surface. The $h\leq 1$ piece of the equation of state has three free parameters, which can be used to enforce continuity and differentiability across $h=1$. In this work we will focus on the choice $a^\prime = n^\prime = 1$ and $K^\prime$ being used to enforce only continuity of $\rho(h)$ across $h=1$. Continuity gives $K^\prime = (1/2) (K(1+n)/a)^n$, then the exterior equation of state is $\rho(h) = h/(2 K^\prime)$. This atmosphere equation of state corresponds to a stiff fluid with  sound speed equal to the speed of light in vacuum, $c_{\rm s} = 1$. The pressure is determined by the indefinite integral
\begin{eqnarray}
p(h) = \int \rho  (h)dh = \left\{ {\begin{array}{*{20}{l}}
{K{\rho ^\Gamma } - \frac{1}{{4{K^\prime }}},}&{h > 1}\\
{{K^\prime }{\rho ^2} - \frac{1}{{4{K^\prime }}},}&{h \le 1}
\end{array}} \right.,
\label{eq:genpoly2}
\end{eqnarray}
where $\Gamma = 1 + 1/n$. The integration constant, which amounts to a cosmological constant, was fixed  by enforcing $p(1) = 0$, that is, the pressure vanishes on the stellar surface, $h=1$. Finally, we choose $a$ according to a specified value of rest mass density at the stellar surface. Let this rest mass value be $\rho_{1}$, then $a=K(1+n) \rho_{1}^{1/n}$.

This equation of state supports $h<1$, which allows us to initialize our simulations with the star and atmosphere both in equilibrium. We initialize $h$ via $H=\alpha h=$ constant $=  \alpha (r\! =\! 0) h(r\! =\! 0)\equiv H_0$, i.e.~$h(r) = H_0/\alpha(r)$. This  specific enthalpy smoothly crosses $h=1$ at the stellar surface, becoming $<1$ outside the star. This smooth behavior of the specific enthalpy makes it a natural choice of a reconstruction variable (instead of $\rho$). Thus, when using the equilibrium atmosphere, we choose to reconstruct $h$ and $\nu^r$ at the cell interfaces.

The enforced rest mass density floor in this case varies in space. We use
\begin{eqnarray}
{\rho _{{\rm{floor}}}}(r) = \left\{ {\begin{array}{*{20}{l}}
{\rho {|_{h = 1}},}&{r < {R_*}}\\
{\rho (r){|_{t = 0}},}&{r \ge {R_*}}
\end{array}} \right.,
\end{eqnarray}
where $R_{*}$ is the stellar surface. If $\rho(r)$ (or $D(r)$) becomes less than $\rho_{\mathrm{floor}}(r)$, then we reset $\rho(r) = \rho_{\mathrm{floor}}(r)$ (or $D(r) = \rho_{\mathrm{floor}}$) and $\nu^r(r) = 0$ (or $p_r=0$). We choose $\rho|_{h=1}=10^{-13}$ in code units, which should be compared with the stellar central density $\rho_c = 1.28 \times 10^{-3}$. The outer boundary at $r=12$ in code units corresponds to $\sim 1.26 R_{*}$. We evolve the star for 10 ms at spatial resolutions $dr = \left\{ 0.2,0.1,0.05,0.025 \right\}$ in code units (or $\left\{0.296,0.148,0.074,0.037\right\}$ km). Since we wish to observe the spatial convergence, we use a fixed time step across these resolutions. This is achieved by using the corresponding CFL factors $\left\{ 0.11875, 0.2375, 0.475, 0.95 \right\} $.

 We note that the fiducial atmosphere is meant to be used only in the hydrodynamic sector of a numerical code. In the gravitational sector, a `mask' must be applied to the energy-momentum tensor $T_{\mu \nu}$ before solving the Einstein equations, such that  $T_{\mu \nu}=0$ when $h<1$. In this work, we use the Cowling approximation, so no mask is required.

Most  neutron star simulation codes readily implement piecewise polytropes as a barotropic or `cold'  EOS to approximate candidate neutron star equations of state. A `hot' EOS term is often added (or, less often,  the polytropic constants $K$ are made temperature-dependent) in order to obtain a baroclinic EOS. In all of the above cases,  the outermost piece(s) of the cold polytrope can be replaced with the generalized  polytrope given by
Eqs.~(\ref{eq:genpoly}-\ref{eq:genpoly2}).  
A parametrization that uses generalized piecewise polytropes with continuous, strictly  positive sound speed for all pieces is part of an upcoming paper 
\cite{oboyleParametrizedEquationState2020}.
Hence, the regularization scheme described above is applicable to a wide class of 
barotropic or baroclinic equations of state. Moreover, a typical nuclear EOS, such as SLy4, has approximately constant sound speed in the outer $\sim 1~\rm km$ of a neutron star, which is beneficial for hyperbolicity. One can extrapolate this constant sound speed out to the surface, or match to the equation of state of iron, keeping the temperature and sound-speed finite on the surface
\cite{haenselAnalyticalRepresentationsUnified2004}. 
\section{Results}

\subsection{Valencia vs hybrid formulation}

In this comparison, we evolve an equilibrium TOV star with a fixed spacetime (Cowling approximation). The initial central density of the star is 
$\rho_{\rm c} = 1.28\times 10^{-3}$, and we use a polytropic equation of state $P = K \rho^\Gamma$ with $\Gamma=2$ and $K=100$. This is a simplistic model of a cold neutron star with gravitational mass $1.4$ M$_{\odot}$, baryonic mass $1.5$ M$_{\odot}$, and radius $R_{*} \approx 14.15$ km. For the hybrid scheme, Valencia is used at the last 2 interior stellar points and all points exterior to those, i.e.~$i_{\mathrm{mix}} = i_{\mathrm{surface}}-1$.

\begin{figure}[t]
%\centering
\centerline{\includegraphics[width=1\textwidth]{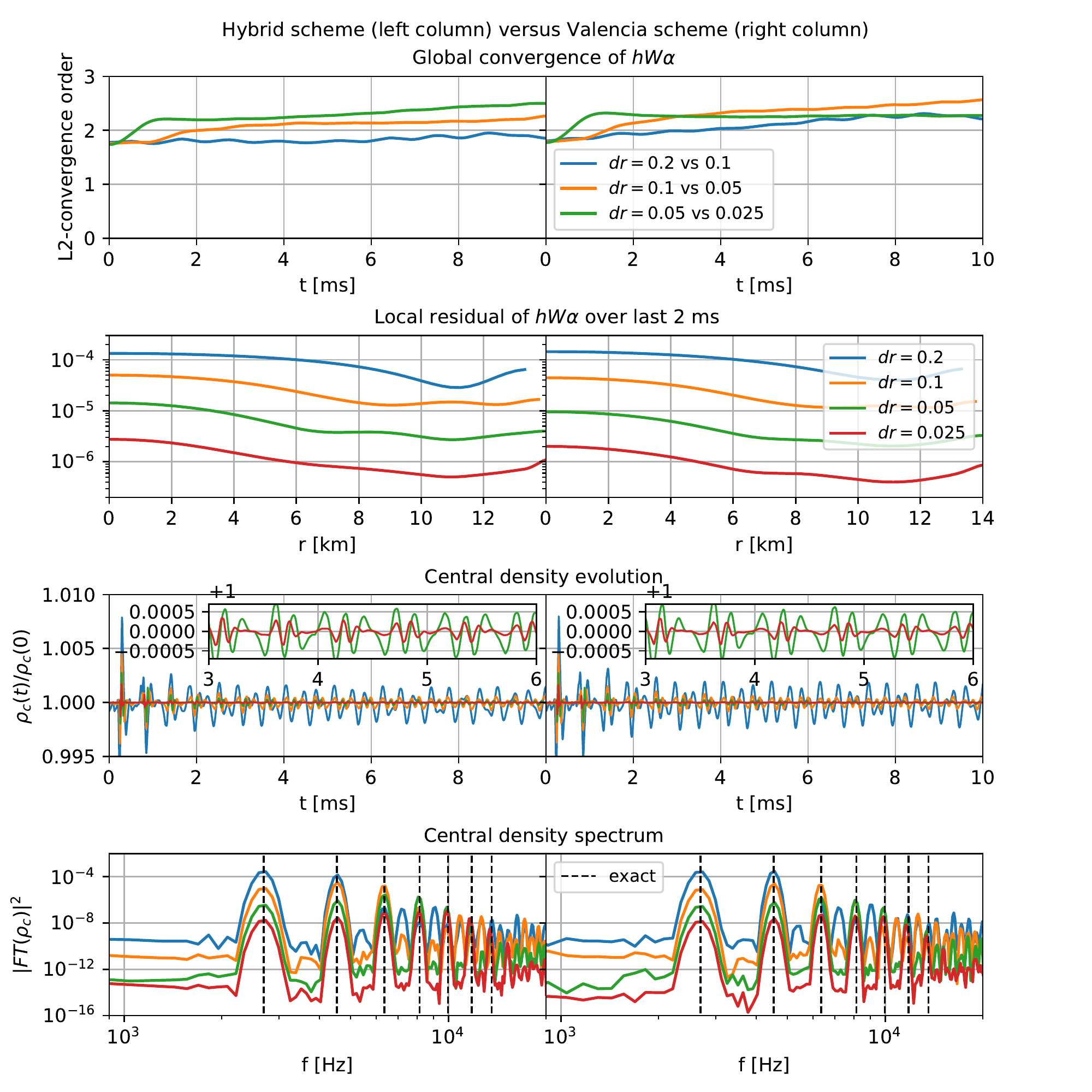}}

\caption{ A comparison between the Valencia formulation (right column) and the hybrid formulation (left column) with $i_{\mathrm{mix}}=i_{\mathrm{surface}}-1$. We evolve an equilibrium TOV star with central rest mass density $\rho(r=0,t=0) \equiv \rho_{\rm c}(t=0) = 1.28\times 10^{-3}$ and equation of state $P=100 \rho^2$ for a total time $T=2027$ in code units. With $G=c=M_{\odot}=1$, this corresponds to a star with gravitational mass $1.4 M_{\odot}$, rest mass $\sim 1.5 M_{\odot}$, and radius $\sim 14.15$ km \cite{fontAxisymmetricModesRotating2001} evolved for $\sim 10$ ms. The time step is kept fixed across all resolutions in order to isolate the spatial error. The Courant factor in the highest resolved case is $0.95$. Our choice of $i_{\mathrm{mix}}=i_{\mathrm{surface}}-1$ is quite extreme (i.e. close to the surface); choosing $i_{\mathrm{mix}}$ instead such that $r_{i_{\mathrm{mix}}} = 13$ km (i.e. using the Hamiltonian formulation in the interior $\sim 92\%$ of the star by areal radius) yields evolution that's almost indistinguishable from Valencia.} \label{fig:hyb_vs_val}
\end{figure}

The results are displayed in Fig.~\ref{fig:hyb_vs_val}. The left column displays the hybrid scheme, and the right column displays the Valencia scheme. In the first row we display the global $L_2$-convergence of the Hamiltonian $H=W\alpha h$ over time. Both schemes give a similar convergence order of $\sim 2$. In the second row we display the local residual of $H$ averaged over the last 2 ms of the evolution. In the third row we display the normalized central rest mass density over time. In the fourth row we display the frequency spectrum of the central density oscillations. To generate these spectra, we apply a Gaussian window $\exp{[-(t-5\mathrm{ ms})^2/(2 \sigma^2)]}$ of width $\sigma=1.2$ ms to $\rho_{\rm c}(t)/\rho_{\rm c}(0)-1$ before computing the Fourier transform. In all of these comparisons, both schemes produce similar results.

%NEWRYAN
Since the equilibrium flux $H$ is constant in the Hamilton-Jacobi conservation law~(\ref{eq:HJgrad31}), one may have expected instead that the hybrid formulation would preserve the equilibrium configuration of the star to a greater degree than the Valencia formulation. In the Valencia formulation, the pressure gradient and source terms in the Euler equation must balance in hydrostatic equilibrium, but since they are discretized differently they do not balance at the numerical level. On the other hand, the Hamilton-Jacobi conservation law~(\ref{eq:HJgrad31}) is balanced without any source terms. In our experimentation we found that the use of the HLL flux is chiefly responsible for the failure of this expectation. Notwithstanding the perturbations injected into the star from the surface, using simple finite differences for the Hamilton-Jacobi flux preserves the stellar equilibrium to a much greater degree than using the finite volume scheme with the HLL flux formula. In light of this, we will be exploring more optimized numerical approaches for the Hamiltonian formulation in future work.

\subsection{Standard vs equilibrium atmosphere}
In this section we explore an alternative vacuum regularization consisting of a modified equation of state which extends to $h<1$, allowing the artificial atmosphere to have a constant Hamiltonian coinciding initially with the Hamiltonian inside the star. The atmosphere and star are therefore initially in equilibrium.

\begin{figure}[t]
%\centering
\centerline{
\includegraphics[width=1\textwidth]{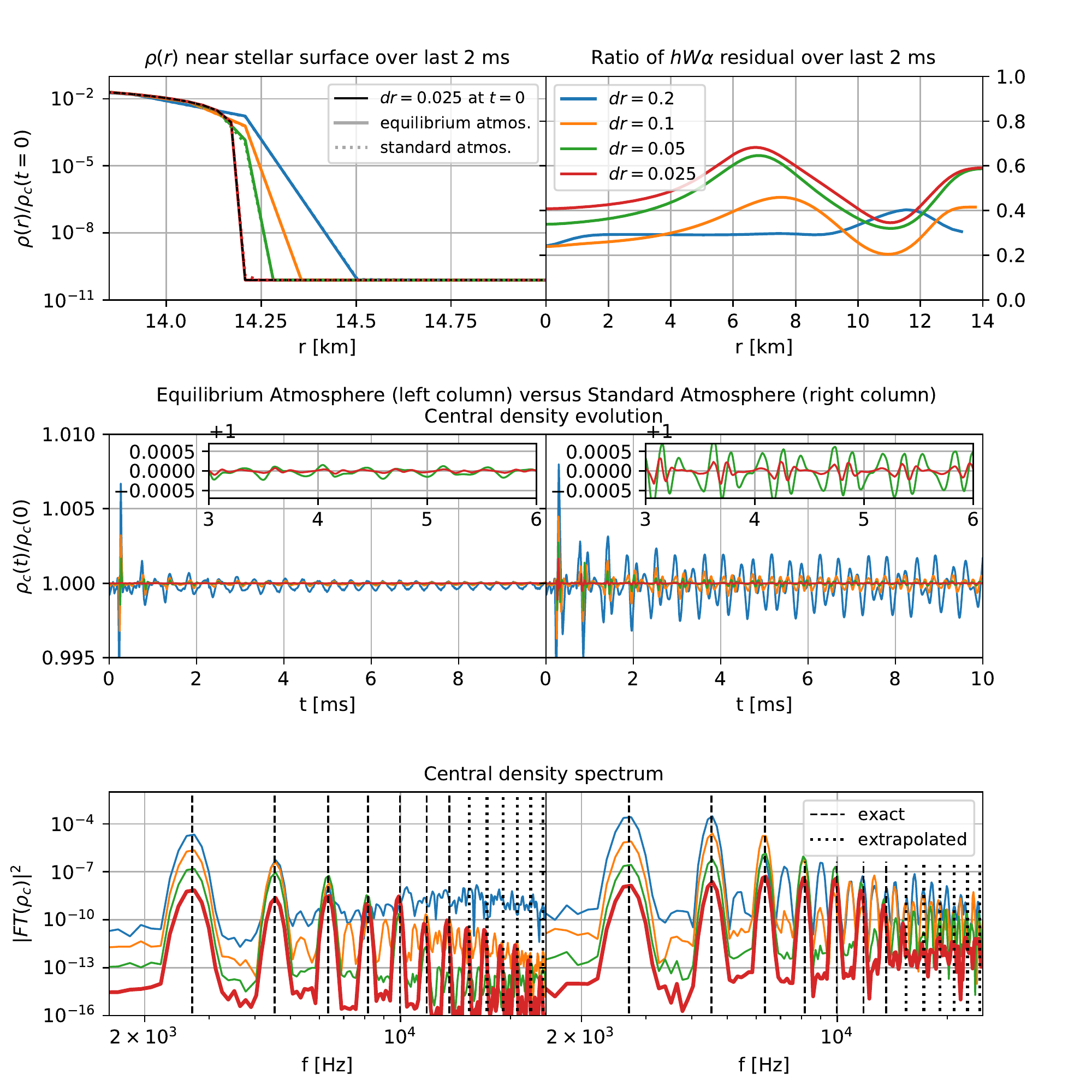}
}

\caption{ A comparison between the standard atmosphere and equilibrium atmosphere for the Valencia formulation. (Top Left): The rest mass density is displayed on log scale near the stellar surface, where the equilibrium atmosphere (solid lines) is found to have an equally sharp profile as the standard atmosphere (dotted lines). (Top Right): Ratios of the local residual of $W\alpha h$ averaged over the last 2 ms, comparing the equilibrium and standard atmosphere treatments. The equilibrium atmosphere treatment yields a $\sim 70 \%$ lower error at low resolution, with this advantage diminishing to $\sim 50 \%$ as resolution is increased. (Middle and Bottom Panels): Similar to Fig.~\ref{fig:hyb_vs_val}, comparing the equilibrium atmosphere (left column) to the standard atmosphere (right column). The central density oscillations are significantly smaller for the equilibrium atmosphere treatment, and the stellar oscillation frequencies are captured correctly. At high resolution (red), the central density oscillation spectra appear to have more modes resolved. We do not have exact frequencies in those cases, so we linearly extrapolate them using the first 7 mode frequencies, obtaining $\{ 15431, 17248, 19066, 20884, 22702, 24519 \}$ Hz. The time step is fixed in all runs, with a Courant factor of $0.95$ for $dr=0.025$.} \label{fig:vacreg}
\end{figure}

In Fig.~\ref{fig:vacreg} we compare 10 ms evolutions of the stationary star using the standard atmosphere and the equilibrium atmosphere. No explicit perturbations are added to the star, thus the oscillations are excited by numerical truncation error. The top left panel shows that the stellar surface is equally sharp around the final 2 ms for both atmospheres. The top right panel shows the ratio of the local Hamiltonian residual $W\alpha h- (W\alpha h)\vert_{t=0}$ averaged over the final 2 ms, with the equilibrium atmosphere in the denominator. The residual in the equilibrium atmosphere case is $\sim 70\%$ smaller than the standard atmosphere case at lower resolutions, with the benefit reducing to $\sim 50\%$ at higher resolution. Note that the lowest resolution has 47 points across one radius of the star, which is a similar resolution to that typically used in binary neutron star simulations. The middle panel compares the central density evolutions, with the equilibrium atmosphere exhibiting much smaller oscillations for all resolutions. The bottom panel compares the central density oscillation spectra. In the highest resolved case (red), the equilbrium atmosphere appears to resolve several more high frequency modes than the standard atmosphere case. Although we do not have exact mode frequencies to compare to for those high overtones, we linearly extrapolate them using the first 7 mode frequencies. The extrapolated frequencies are $\{ 15431, 17248, 19066, 20884, 22702, 24519 \}$ Hz, rounded to the nearest 1 Hz. The linear fit of the first 7 mode frequencies has an $\mathcal{L}_2$-norm disagreement with those frequencies of $\sim 8$ Hz.

\begin{figure}[t]
%\centering
\centerline{
\includegraphics[width=0.5\textwidth]{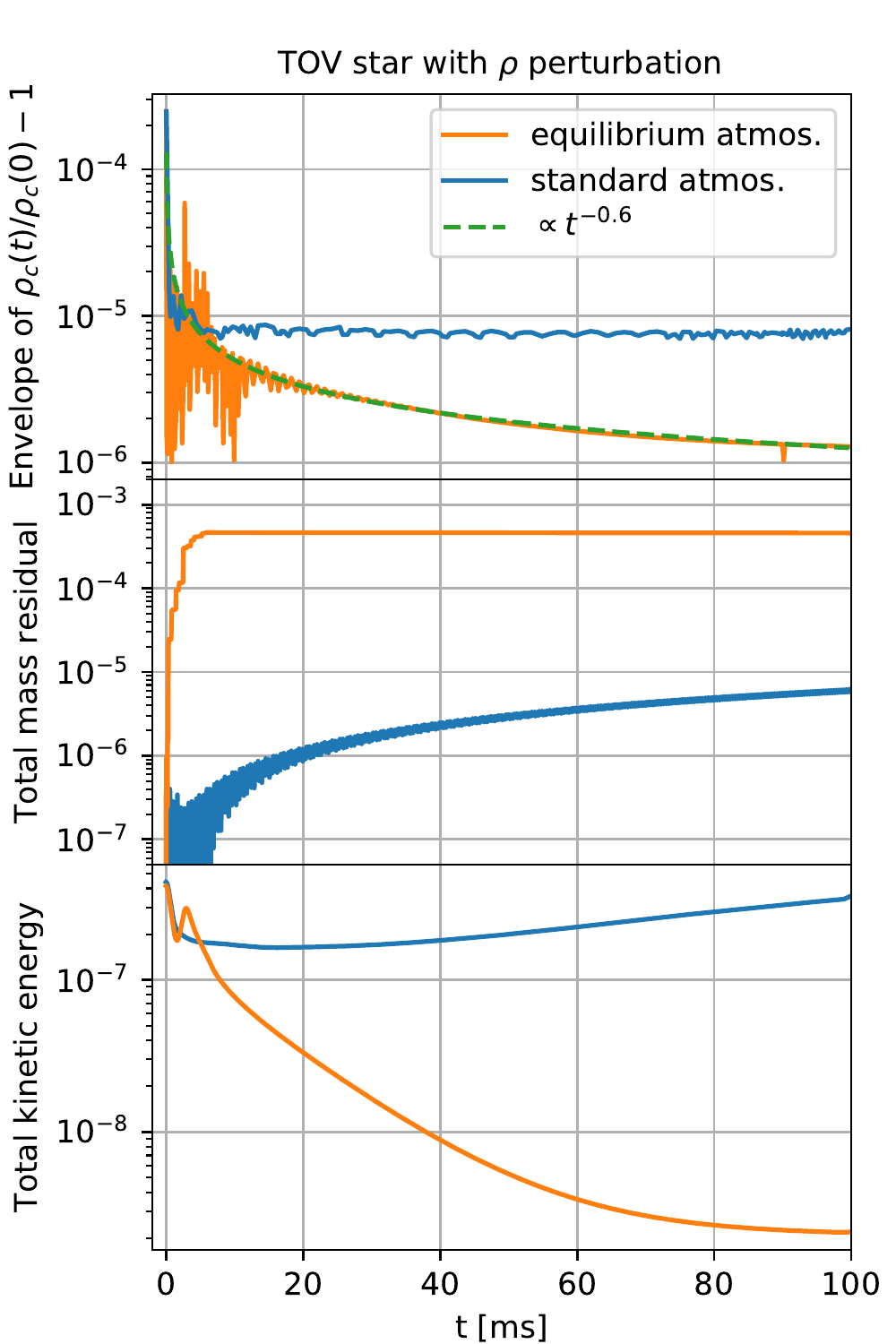}
\includegraphics[width=0.5\textwidth]{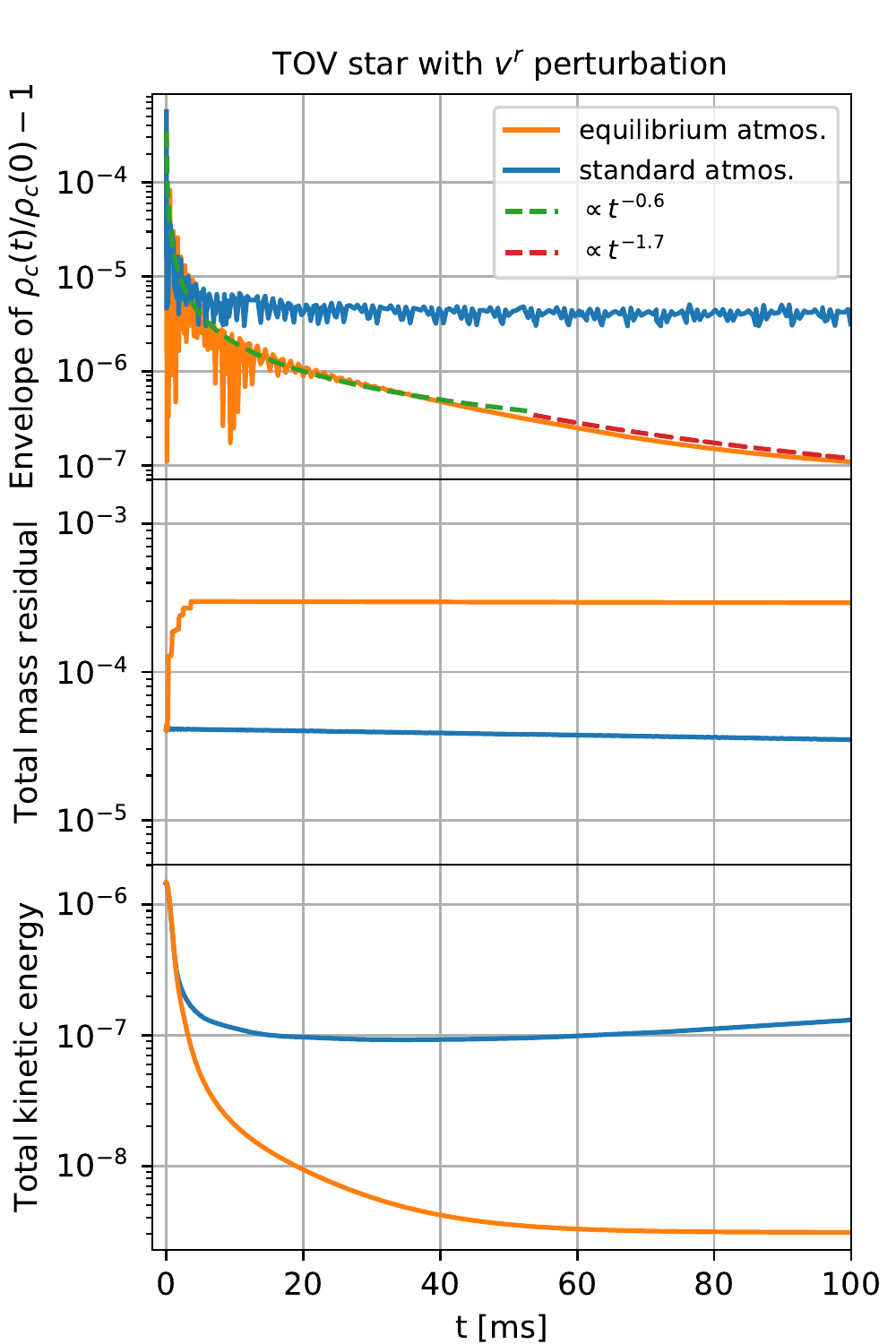}
}
%[width=1\textwidth]

\caption{ A comparison between the standard atmosphere and equilibrium atmosphere for a TOV star initialized with a density perturbation (left) and a velocity perturbation (right), in the Valencia formulation. (Top panel): The oscillation amplitude of the central rest mass density residual over time, computed by extracting the local maxima in time. The equilibrium atmosphere treatment results in a $\sim t^{-0.6}$ decay of the oscillations, steepening in the case of a velocity perturbation to $\sim t^{-1.7}$ around 50 ms. (Middle panel): The total relativistic rest mass on the grid over time. On the left, the equilibrium atmosphere treatment rapidly sheds much of the initial excess mass, whereas the standard atmosphere treatment retains it for a longer period of time. On the right there is also 10x more mass loss for the equilibrium atmosphere. (Bottom panel): The total kinetic energy over time. The kinetic energy density is taken to be $\rho h (W^2-1)$. The equilibrium atmosphere produces decay, whereas the standard atmosphere produces gradual growth. This suggests that the more sustained central density oscillations observed in the top panel for the standard atmosphere are due to the injection of kinetic energy into the star by the vacuum regularization treatment at the surface.} \label{fig:vacreg_pert}
\end{figure}

In Fig.~\ref{fig:vacreg_pert} we compare 100 ms evolutions at low-moderate $dx=0.1$ resolution initialized with explicit density (left column) or velocity (right column) perturbations. There are $\sim 94$ points across one radius of the star at this resolution. The initial perturbations are given in code units by
\begin{eqnarray}
\delta \rho(r) &=& 0.05 \rho_c \exp{\left[-\frac{(r-3)^2}{2*0.2^2}\right]} \\
\delta v^r(r) &=& 0.05 \exp{\left[-\frac{(r-3)^2}{2*0.2^2}\right]}.
\end{eqnarray}

Note that $R_{*} \sim 9.56$ in code units. These perturbations are rather extreme in comparison to those expected in the inspiral phase of a NSNS or BHNS binary. However, our main purpose here is to compare the stability of the equilibrium and standard atmosphere treatments.
 In the top row we plot the amplitude envelope of the central density fluctuations. The standard atmosphere roughly maintains a certain amplitude of fluctuations, whereas the equilibrium atmosphere treatment yields decaying oscillations with $\sim{t^{-0.6}}$ behavior for both types of perturbation, hastening to $\sim{t^{-1.7}}$ after 50 ms in the case of a velocity perturbation. 
  The middle panel displays the total relativistic rest mass residual over time. The equilibrium atmosphere does significantly worse, $\sim{40}\times$  for the density perturbation and $\sim8\times$ so for the velocity perturbation, although remaining below the $10^{-3}$ level. Given the extreme amplitude of these perturbations, this may not be a significant issue during the inspiral phase in future applications, although that remains to be seen.
The bottom panel shows the total relativistic kinetic energy over time, which we define as $\rho h (W^2-1)$. The standard atmosphere treatment settles to a much larger value than the equilibrium atmosphere, before increasing moderately. This indicates that the standard atmosphere injects much larger perturbations into the star than the equilibrium atmosphere. The kinetic energy in the equilibrium atmosphere case decays and settles down to a level $\sim{100}\times$ less than the standard atmosphere case. For comparison, in the stationary star evolutions at this same resolution, the total mass decays for both atmosphere treatments but is preserved at the $10^{-7}$ level. Therefore we expect the mass preservation in a binary simulation would be comparable for both atmosphere treatments, with the main difference being the reduced amplitude of fluctuations when using the equilibrium atmosphere treatment.

Our intent is to illustrate this equilibrium atmosphere as an initial exploration of an alternative vacuum regularization method. Our results show some promising features, namely the greater preservation of the equilibrium star and the resolution of more overtone modes compared to the standard atmosphere treatment. On the other hand, the decay of perturbations in spherical symmetry is not necessarily desired if the perturbations are physically sourced, because the only physical mechanism to damp them would be viscosity. Thus, in our results the equilibrium atmosphere exhibits an artificially high level of dissipation. Further exploration of these ideas, together with more optimized numerical approaches for the Hamilton-Jacobi conservation law~(\ref{eq:HJgrad31}), will be explored in future work.

\section{Summary}
Hydrodynamic simulations in numerical general relativity  typically employ the Valencia scheme in combination with a shock-capturing discretization method. During the inspiral phase of binary neutron star evolution, the flow can be well-modelled as barotropic and shocks are absent, so  Kelvin's theorem holds. Most simulations start with irrotational initial data, which is considered a good approximation tens of orbits before merger, when the orbital frequency is much higher than the spin frequency. Then, Kelvin's theorem guarantees that no canonical vorticity  will develop during inspiral. Since the flow remains irrotational and barotropic until tidal disruption and merger, one may use a Hamilton-Jacobi formulation to simulate the inspiral phase.

Towards this goal, we presented a first implementation of Hamilton-Jacobi hydrodynamics in numerical relativity, focusing here on radially pulsating neutron stars. Obtaining stable, convergent evolutions with the correct mode frequencies is a nontrivial test, as it requires a well-posed (strongly hyperbolic) formulation, with suitable treatment of the vacuum boundary. In particular, for a perfect fluid (no fluid viscosity) in spherical symmetry (no gravitational wave dissipation), perturbations traveling outward must be completely reflected inward at the free surface in order to obtain the correct mode frequencies. We further demonstrated 
that a simple regularization of the Euler equations on the vacuum boundary, that preserves hyperbolicity, can be accomplished by preserving sound-speed positivity at the EOS level. This allows the pressure (but not the density) to be exactly zero on the stellar surface in hydrodynamic simulations, which has not been possible before.
Notwithstanding the possible relevance of a physical low-density atmosphere around neutron stars, an artificial one is no longer required for preserving hyperbolicity at the surface.
Nevertheless, in this work, we opted to still use a (fiducial) equilibrium atmosphere, in order to obtain reflections at the vacuum boundary and obtain the correct frequency modes. We found that the equilibrium atmosphere is dissipative and less noisy and allows one to compute higher frequency modes than the standard atmosphere treatment.
Kastaun~\cite{kastaunHighresolutionShockCapturing2006} developed a flux-balanced scheme (where the Valencia source terms are recast as fluxes, using \emph{a priori} information about the equilibrium solutions of stars). This scheme, combined with an alternative surface treatment, was also found to be dissipative.

Although a shock-capturing method is in principle not necessary during the inspiral, current binary neutron star simulation codes, with the standard atmosphere treatment, see the stellar surface   as a shock discontinuity, and shock-capturing schemes appear to be needed in order to obtain the correct mode frequencies. We found that, although the Hamilton-Jacobi formulation can be used with the HLL scheme, errors are higher near the surface compared to the Valencia formulation. Alternative discretization tests,  comparisons with well-balanced schemes \cite{levequeBalancingSourceTerms1998,kappeliWellbalancedSchemesEuler2014} and improved vacuum boundary treatment \cite{toroRiemannSolversNumerical2009,tsakirisVacuumTrackingMethod2007}
are the subject of future work.

In the current scheme, perturbations that travel outward are partially reflected inward and partially transmitted outside when they reach the stellar surface ($h=1$). The vacuum variables are reset only if they drop below the floor, but are allowed to evolve otherwise. Thus, some oscillatory energy escapes into the fiducial atmosphere, which is responsible for the observed dissipation of radial oscillations. If this dissipative treatment is used and compared in a quasi-circular inspiral to a standard non-dissipative treatment, 
one could potentially disentangle the gravitational-wave phase accumulation due to tidal effects from the phase accumulation due to (tidally excited) f-mode oscillations during inspiral \cite{schmidtFrequencyDomainModel2019}. Measuring the phase accumulation due to these two effects separately would enable better comparison and calibration with semi-analytical gravitational wave models of binary neutron-star inspiral. 

If one evolved the velocity potential $S$ directly, via the Hamilton-Jacobi equation, the flow would be numerically guaranteed to remain irrotational, satisfying the theorems of Kelvin and Helmholtz exactly. 
In this work, we instead evolve the gradient of the Hamilton-Jacobi equation, which has a hyperbolic flux-conservative form, allowing use of standard shock-capturing schemes. However, evolving the gradient of the Hamilton-Jacobi equation means that the irrotationality constraints~(\ref{eq:canmomirr}) and (\ref{eq:HJconstraint}) may be numerically violated. In a companion paper we will explore the use of constraint damping to alleviate this problem, ensuring 
that the third Helmholtz theorem is numerically satisfied. This constraint damping scheme is feasible within the Hamilton-Jacobi formulation but not the Valencia formulation.

While the Hamiltonian and Valencia formulations are equivalent for differentiable solutions, this need not be the case for weak  solutions. The fact that
the stellar surface is treated as a ``shock'' discontinuity by the HLL scheme is related to the issues  near the surface  that we faced with the Hamiltonian formulation, which we remedied using a hybrid scheme near the surface~\cite{JRWSbaroclinic}.
Finally, this work focused on barotropic flows, which are applicable
throughout the inspiral phase of a binary. During tidal discruption and merger, shock heating generates entropy, and the barotropic Hamiltonian formulation no longer applies. In a companion paper~\cite{JRWSbaroclinic}, baroclinic Hamiltonian  formulations are explored and found  to admit unphysical shock solutions.  Thus, pending a possible innovative remedy, it is advisable to use an explicitly barotropic Hamiltonian formulation during the inspiral phase, and then switch to the baroclinic Valencia formulation just prior to merger.

\section*{Acknowledgments}
We thank Nils A. Andersson, Brandon Carter, 
Kyriaki Dionysopoulou, John L. Friedman, Charles F. Gammie, Eric Gourgoulhon, Carsten Gundlach, Roland Haas,  David Hilditch, Ian Hawke, Christian Kr\"uger,  Koutaru Kyutokou, Niclas Moldenhauer, Vasilis Paschalidis, Frans Pretorius, David Radice, 
Luciano Rezzolla, Stuart Shapiro, Ulrich Sperhake, Masaru Shibata, Nikolaos Stergioulas and Yifan Zhang for valuable input.
 This work was supported by  the European Union’s Horizon 2020 research and innovation programme under the Marie Sk\l odowska-Curie grant agreement No 753115 and COST Action NewCompStar, by the National Science Foundation Grant
PHY-1912619 at the University of Arizona and by NCSA's Students Pushing Innovation (SPIN) program. A subset of computations were performed on the Ocelote cluster at the University of Arizona.
The authors thank the Isaac Newton Institute of Mathematical Sciences for  hospitality during the program ``Structure preservation and general relativity'' under EPSRC Grant Number EP/R014604/1.
Part of this work was completed  during the second author's secondments at the Aristotle University of Thessaloniki and Princeton University. CM thanks the AUTh and Princeton relativity groups for their stimulating research environment and warm hospitality.
\section*{References}

\bibliographystyle{iopart-num}
\bibliography{Arizona}

%http://127.0.0.1:23119/better-bibtex/collection?/4/2XKVLIBF.biblatex

\appendix
\section{Valencia formulation} \label{app:valencia}

For completeness, in this section we provide the details of the Valencia formulation, in particular the equations of motion in spherical symmetry and the corresponding characteristic structure and discretization. Explicit formulae for the characteristic structure in the Valencia formulation are well-known, but for modularity of our numerical implementation we instead solve for the eigenstructure numerically at each point, in the same way as when using the Hamiltonian formulation.

The conservative variables are $\vec{U} = (\tilde{D}, \tilde{S}_r) = (\Upsilon \rho W, \Upsilon \rho h W^2 \gamma_{rr} \nu^r)$. The equations of motion read
\begin{eqnarray}
0 &=& \partial_t \tilde{D} + \frac{1}{r^2} \partial_r \left[ \alpha r^2 \tilde{D} \left( \nu^r - \beta^2/\alpha \right)\right] \\
0 &=& \partial_t \tilde{S}_r + \frac{1}{r^2} \partial_r \left[ \alpha r^2 \tilde{S}_r \left(\nu^r - \beta^r/\alpha \right) \right] + \partial_r \left[ \alpha \Upsilon P \right] - \mathcal{S},
\end{eqnarray}
where the source term $\mathcal{S}$ is given by
\begin{eqnarray}
 \mathcal{S} &=& \alpha \Bigl[ \partial_r \ln{\gamma_T} P \Upsilon - \left(\partial_r \ln{\alpha}\right) \left( \tilde{\tau} + \tilde{D} \right)  \nonumber\\
 &+& \frac{1}{2} \left(\partial_r \ln{\gamma_{rr}} \right) \left( \tilde{S}_r \nu^r + P \Upsilon \right) + \frac{\left(\partial_r \beta^r\right)}{\alpha} \tilde{S}_r \Bigr],
\end{eqnarray}
where we have defined $\tilde{\tau} = \Upsilon \left(\rho h W^2 - P - \rho W \right)$.

For the purposes of characteristic analysis, the fluxes are $\vec{F} = (F_{\tilde{D}},F_{\tilde{S}_r} + \alpha \Upsilon P) = (\alpha \tilde{D} (\nu^r-\beta^r/\alpha), \alpha \tilde{S}_r (\nu ^r-\beta^r/\alpha) + \alpha \Upsilon P)$. Then the matrices analogous to Eqs.~(\ref{eq:dUdq}) \&~(\ref{eq:dFdq}) are
\begin{eqnarray}
\frac{\partial \vec{U}}{\partial \vec{q}}
&=& 
\left[ {\begin{array}{*{20}{c}}
W&{\Upsilon \rho {W^3}{\gamma _{rr}}{\nu^r}}\\
{(h + \rho {\partial _\rho }h)\Upsilon {W^2}{\gamma _{rr}}{\nu^r}}&{\Upsilon \rho h{\gamma _{rr}}{W^2}(1 + 2{W^2}{\gamma _{rr}}{\nu^r}{\nu ^r})}
\end{array}} \right]
\label{eq:dUdq_val} \\
\frac{\partial \vec{F}}{\partial \vec{q}} 
&=&
\left[ {\begin{array}{*{20}{c}}
{\alpha W({\nu^r} - {\beta ^r}/\alpha )}&{\Upsilon \alpha \rho W[1 + {W^2}{\gamma _{rr}}{\nu^r}({\nu^r} - {\beta ^r}/\alpha )]}\\
{{\partial _\rho }{F_{{{\tilde S}_r}}} + \alpha \Upsilon {\partial _\rho }P}&{{\partial _{{\nu^r}}}{F_{{{\tilde S}_r}}}}
\end{array}} \right]
 \label{eq:dFdq_val}
\end{eqnarray}
with
\begin{eqnarray}
\partial_\rho F_{\tilde{S}_r} &=& \Upsilon \alpha \left(h + \rho \partial_{\rho}h \right) W^2 \gamma_{rr} \nu^r \left(\nu^r - \beta^r/\alpha\right) \nonumber\\
\partial_{\nu^r} F_{\tilde{S}_r} &=& \Upsilon \alpha \rho h \gamma_{rr} W^2 \left[ 2 \nu^r - \beta^r/\alpha + 2 W^2 \gamma_{rr} \nu^r \nu^r \left( \nu^r - \beta^r/\alpha \right) \right].
\end{eqnarray}

The discretized equations of motion are
\begin{eqnarray}
 (\tilde{D})^{n+1}_i  &=& (\tilde{D})^n_i - 3 \frac{\Delta t}{\Delta(r^3)_i} \left[ r_{i+1/2}^2 ({F_{\tilde{D}}})^{n+1/2}_{i+1/2} - r_{i-1/2}^2 ({F_{\tilde{D}}})^{n+1/2}_{i-1/2} \right] \\
 ({\tilde{S}_r})_{i}^{n+1} &=&  ({\tilde{S}_r})_{i}^n - 3 \frac{\Delta t}{\Delta(r^3)_i} \left[ r_{i+1/2}^2 ({F_{\tilde{S}_r}})^{n+1/2}_{i+1/2} - r_{i-1/2}^2 ({F_{\tilde{S}_r}})^{n+1/2}_{i-1/2} \right] \nonumber\\
&\phantom{=}& \phantom{({\tilde{S}_r})_{i}^n} - \frac{\Delta t}{\Delta r} \left[ (\alpha \Upsilon P)_{i+1/2}^{n+1/2} - (\alpha \Upsilon P )_{i-1/2}^{n+1/2}  \right] + \mathcal{S}_i^n,
\end{eqnarray}
where $r_{i\pm 1/2} = r_i \pm \Delta r/2$.

For the fluxes $F_{\tilde{D}}$ and $F_{\tilde{S}_r}$, we use the HLL flux Eq.~(\ref{eq:HLL}), whereas for the split flux we omit the correction term, i.e. we use
\begin{eqnarray}
\alpha \Upsilon P = \frac{s_R (\alpha\Upsilon P)_L - s_L (\alpha\Upsilon P)_R}{s_R-s_L}.
\end{eqnarray}
This ensures that the correction term $\propto U_R - U_L$ is not used twice. The source term $\mathcal{S}_i^n$ is computed using cell-centered values, which is an approximation since it should be a cell-averaged quantity.

\end{document}